 \documentclass[final,5p,times]{elsarticle}

 \usepackage{graphics}

\usepackage{amssymb}

 \usepackage{lineno}
 \usepackage{sidecap}
 
 \newcommand{\etal}{\MakeLowercase{\textit{et al. }}} 

\journal{NIM A  RICAP-2013}

\begin{document}

\begin{frontmatter}



\title{Tunka-Rex: Status and Results of the First Measurements}


\author[ikp]{D.~Kostunin}
\author[isu]{N.M.~Budnev}
\author[isu]{O.A.~Gress}
\author[ikp]{A.~Haungs}
\author[ikp]{R.~Hiller}
\author[ikp]{T.~Huege}
\author[isu]{Y.~Kazarina}
\author[ipe]{M.~Kleifges}
\author[sinp]{A.~Konstantinov}
\author[isu]{E.N.~Konstantinov}
\author[sinp]{E.E.~Korosteleva}
\author[ipe]{O.~Kr\"omer}
\author[sinp]{L.A.~Kuzmichev}
\author[isu]{R.R.~Mirgazov}
\author[isu]{L.~Pankov}
\author[sinp]{V.V.~Prosin}
\author[inr]{G.I.~Rubtsov}
\author[ipe]{C.~R\"uhle}
\author[ikp]{F.G.~Schr\"oder}
\author[isu]{E.~Svetnitsky}
\author[desy]{R.~Wischnewski}
\author[isu]{A.~Zagorodnikov}

\address[ikp]{Institut f\"ur Kernphysik, Karlsruhe Institute of Technology (KIT), Germany}
\address[isu]{Institute of Applied Physics ISU, Irkutsk, Russia}
\address[ipe]{Institut f\"ur Prozessdatenverarbeitung und Elektronik, Karlsruhe Institute of Technology (KIT), Germany}
\address[sinp]{Skobeltsyn Institute of Nuclear Physics MSU, Moscow, Russia}
\address[inr]{Institute for Nuclear Research of the Russian Academy of Sciences, Moscow, Russia}
\address[desy]{DESY, Zeuthen, Germany}

\begin{abstract}
Tunka-Rex is the radio extension of Tunka-133 located in Siberia close to Lake Baikal. 
The latter is a photomultiplier array registering air-Cherenkov light from air showers induced by cosmic-ray particles with initial energies of approximately $10^{16}$ to $10^{18}$ eV.
Tunka-Rex extends this detector with 25 antennas spread over an area of 1 km$^2$.
It is triggered externally by Tunka-133, and detects the radio emission of the same air showers.
The combination of an air-Cherenkov and a radio detector provides a facility for hybrid measurements and cross-calibration between the two techniques.
The main goal of Tunka-Rex is to determine the precision of the reconstruction of air-shower parameters using the radio detection technique.
It started operation in autumn 2012. We present the overall concept of Tunka-Rex, the current status of the array and first analysis results.
\end{abstract}

\begin{keyword}
Tunka-Rex \sep 
Tunka-133 \sep
Tunka \sep
ultra-high energy cosmic rays \sep
extensive air showers \sep
radio detection


\end{keyword}

\end{frontmatter}

 \linenumbers

\section{Introduction}
Since the measurements of cosmic rays have reached energies of the GZK~\cite{Greisen:1966jv,Zatsepin:1966jv}, the main challenge for the physics of ultra-high cosmic rays is to increase the statistics and the measurement quality close to the breakdown of the cosmic ray flux at approximate 60 EeV. To obtain a sufficient statistics we need to build economically reasonable large-area detectors with high duty cycle. The radio detection could be one of the perspective techniques for future investigations of ultra-high energy cosmic rays.

Radio emission from extensive air showers was theoretically predicted~\cite{Askaryan1961,Kahn1966,Castagnoli1969,Hough1973} and first detected~\cite{Jelley,Vernov,Allan1971} about 50~years ago. 
The radio detection techniques became popular in the last decade again, because standard detection methods have reached technological and economical limits: measurements by  surface particle detectors depend on models, whose accuracy is limited at high energies due to extrapolation; optical fluorescence and air-Cherenkov detectors are limited by their duty cycle due to duration of dark nights and weather. 
Thus, a number of modern experiments~\cite{FalckeNature2005,CODALEMA,AERA,LOFAR} aims at obtaining the main properties of extensive air showers, such as arrival direction, energy, shower maximum and primary particle\footnote{The chemical composition of cosmic rays (i.e, the primary particles) can be reconstructed only by indirect methods from air-shower measurements, for example, by combining study of primary energy and shower maximum, which can be obtained by the optical detectors.} using the radio detection technique.
These experiments proved that the radio emission can be detected from air showers with energies above $10^{17}$~eV, with an angular resolution for the arrival direction better than $1^{\circ}$~\cite{SchroederLOPES_ARENA2012}.

The open question is the precision of the reconstruction for primary energy and shower maximum. Up to now, the experiments have given only upper limits for these quantities ($20\,\%$ for the energy and about $100\,$g/cm\textsuperscript{2} for $X_\mathrm{max}$)~\cite{GlaserAERA_ARENA2012,RebaiCODALEMAenergy2012,ApelLOPES_MTD2012,PalmieriLOPES_ICRC2013}. In a very recent report from the experiment LOFAR, it was estimated that a precision of the $X_{\mathrm{max}}$ reconstruction can reach up to $20$ g/cm$^2$~\cite{LOFAR_ICRC2013}. 
This precision would be comparable with the fluorescence technique. The current challenge is to reach a competitive precision with an economic radio array which can be scaled to very large areas.

The main goal of Tunka-Rex, the radio extension of the Tunka observatory for air showers, is to answer this question, i.e., to determine the precision for the reconstruction of the energy and the atmospheric depth of the shower maximum based on the cross-calibration with a air-Cherenkov detector.
For this purpose, Tunka-Rex is built within the Tunka-133 photomultiplier (PMT) array. The latter is measuring the air-Cherenkov light of air showers in the energy range between $10^{16}$ and $10^{18}\,$eV \cite{TunkaRICAP2013,TunkaICRC2013}. 
Data of both detectors are recorded by a shared data-acquisition system, and the radio antennas are triggered by the photomultiplier measurements. 
This setup automatically provides hybrid measurements of the radio and the air-Cherenkov signal, and consequently allows us to perform a cross-calibration of both techniques. 
In particular, we can test the sensitivity of the Tunka-Rex radio measurements for the energy and for $X_\mathrm{max}$ by comparing them to the measurements of the established air-Cherenkov array.

\section{Setup and hardware properties}
\begin{figure}[t!]
\includegraphics[width=1.0\linewidth]{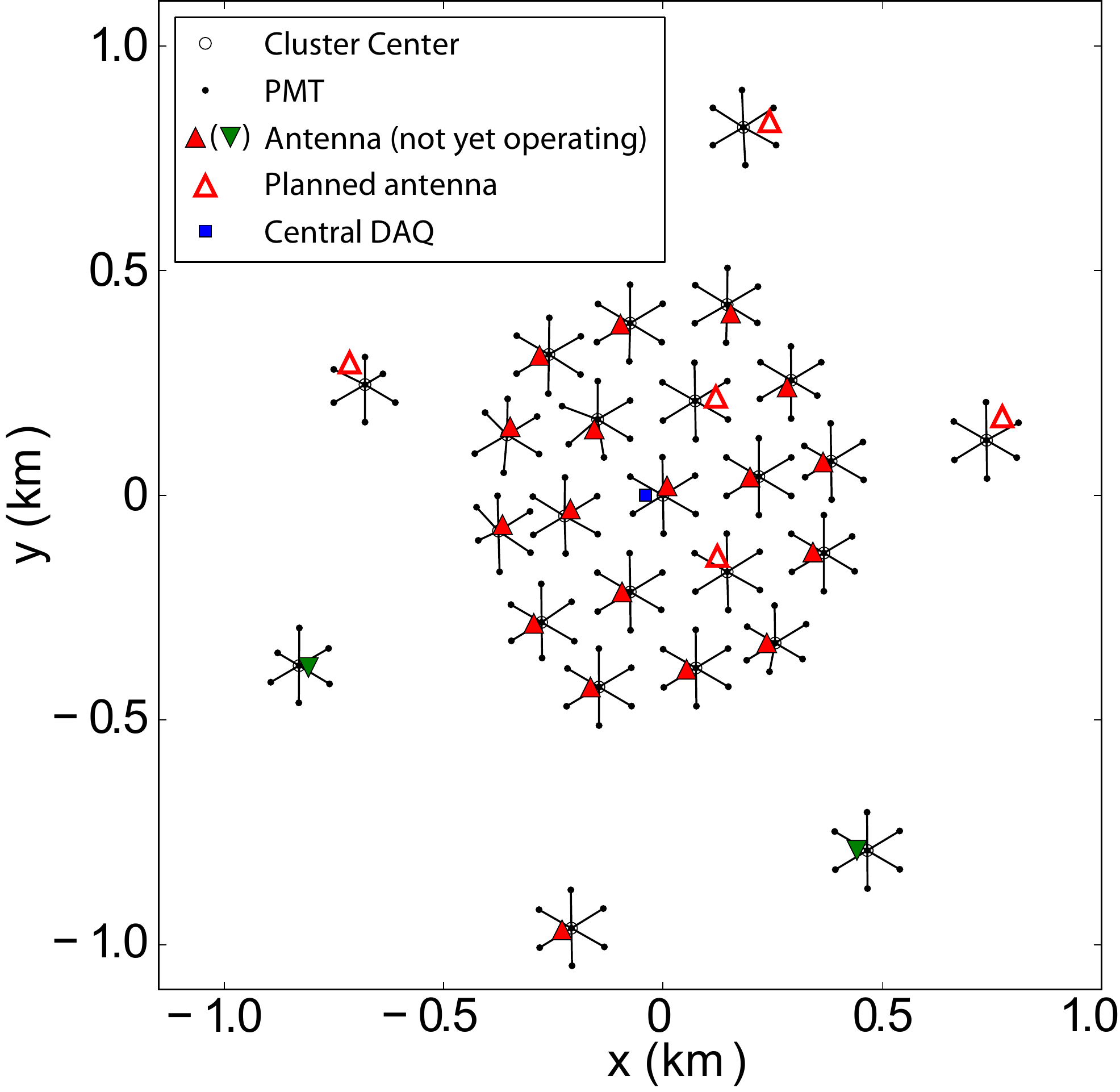}
\caption{Map of Tunka. One Tunka-Rex antenna, consisting of two SALLAs, is attached to each cluster of seven non-imaging photo-multipliers (PMTs). In the season from October 2012 to April 2013, 18 out of 20 Tunka-Rex antennas have been operating. In autumn 2013, five additional antennas will be installed to complete the array of 25 antennas in total.} \label{fig_map}
\end{figure}

Tunka-Rex currently consists of 20 antennas attached by cables to the cluster centers of the Tunka-133 photomultiplier array (Fig.~\ref{fig_map}), which is organized in 25 clusters formed by 7 PMTs each. 

\begin{figure}[h!]
\includegraphics[width=1.0\linewidth]{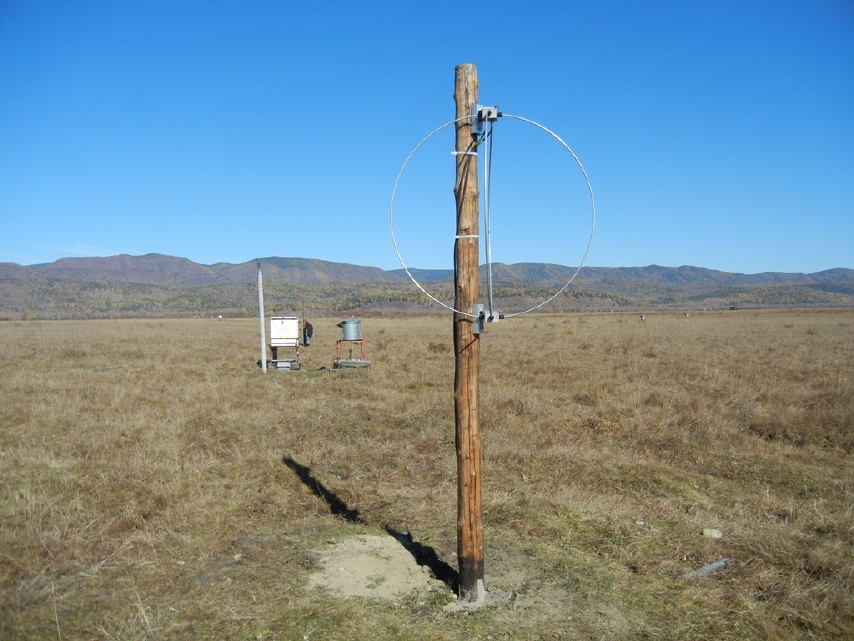}
\caption{A Tunka-Rex antenna in front of a Tunka-133 cluster center box and its central PMT.}
\label{fig-SALLA}
\end{figure}

\begin{figure}[h!]
\centering
\includegraphics[width=1.0\linewidth]{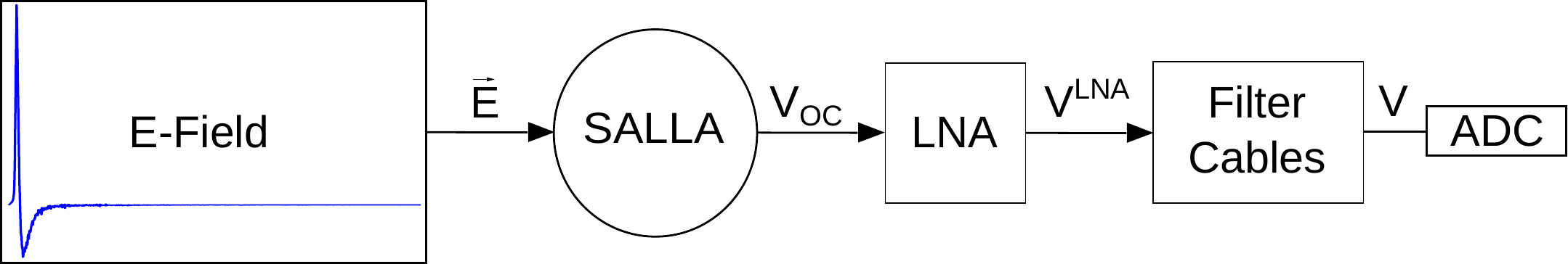}
\caption{Signal chain of a Tunka-Rex antenna with the corresponding transformations.}
\label{chain}
\end{figure}

The spacing between the antennas in the inner clusters is approximately $200\,$m, covering an area of roughly $1\,$km\textsuperscript{2}. 
At each antenna position there are two orthogonally aligned SALLAs (short aperiodic loaded loop antenna) \cite{AERAantennaPaper2012} with 120~cm diameter (Fig.~\ref{fig-SALLA}). 
Unlike most radio experiments the antennas in Tunka-Rex are not aligned along the north-south and east-west axis, but rotated by 45$^{\circ}$, like in LOFAR \cite{LOFAR}. Since the radio signal from cosmic ray air showers is predominantly east-west polarized \cite{polarization}, this should result in more antennas with signal in both channels but also less events with signal in at least one channel.
The SALLA has been chosen as antenna not only for economic reasons, but also because its properties depend only little on environmental conditions, particularly, the influence of the ground on the antenna gain and, thus, the measurement accuracy, is suppressed by a load attached to the bottom part of the antenna. The signal chain is continued by a low noise amplifier (LNA) placed in the isolated metal box, connected directly to the top of the SALLA; a buried coaxial cable connecting the antenna to the main amplifier and a filter at each cluster center of Tunka-133 hosting a flash ADC board for the digital data acquisition (Fig.~\ref{chain}). 

\begin{figure}[h!]
\includegraphics[width=1.0\linewidth]{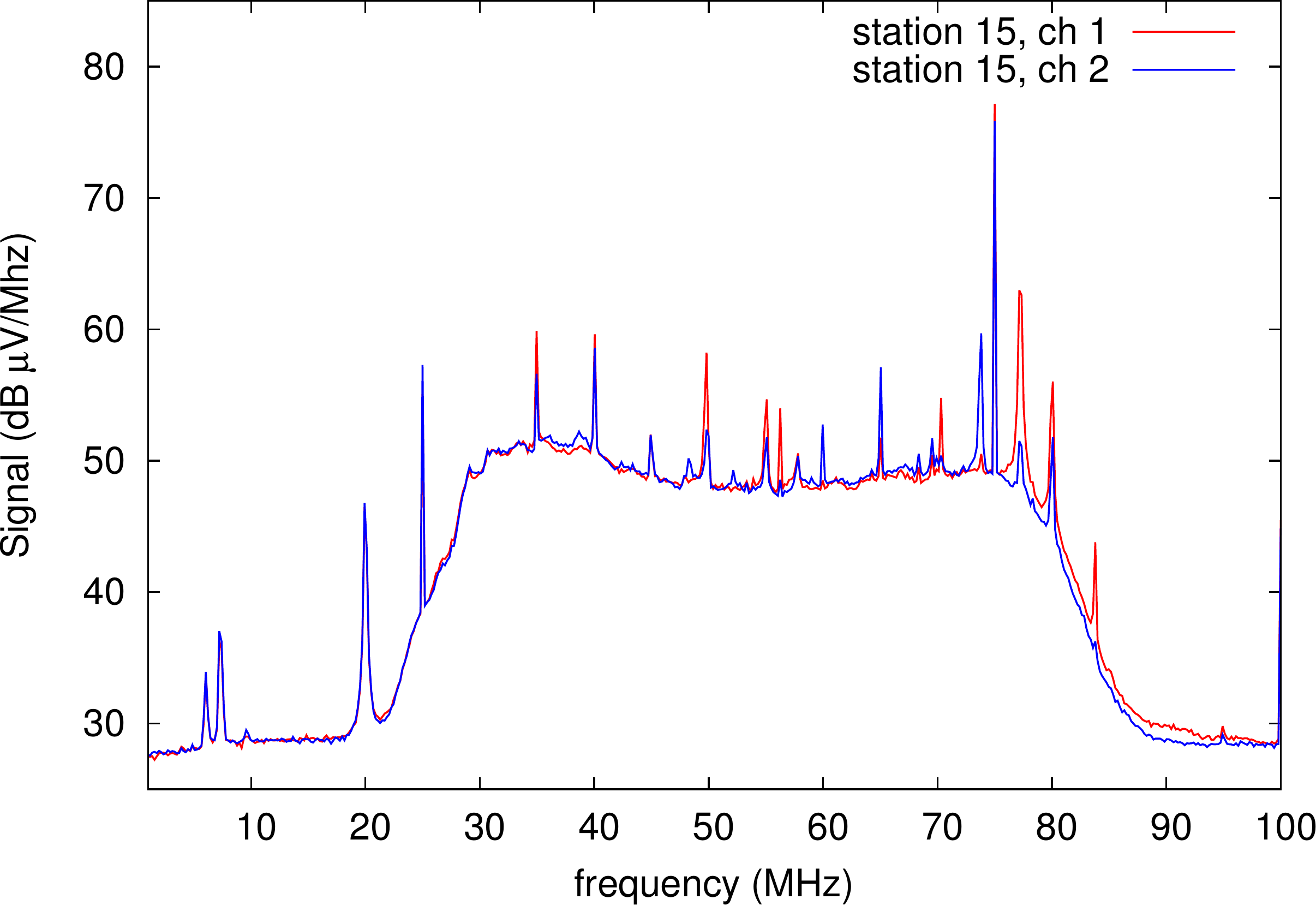}
\caption{The mean background spectrum in one night of operation, as measured with Tunka-Rex station 15. Background interferences are under active investigation, more details see Ref.~\cite{HillerTunkaRex_ICRC2013}} 
\label{fig-spectrum}
\end{figure}
\begin{figure}[t!]
\centering
\includegraphics[width=0.85\linewidth]{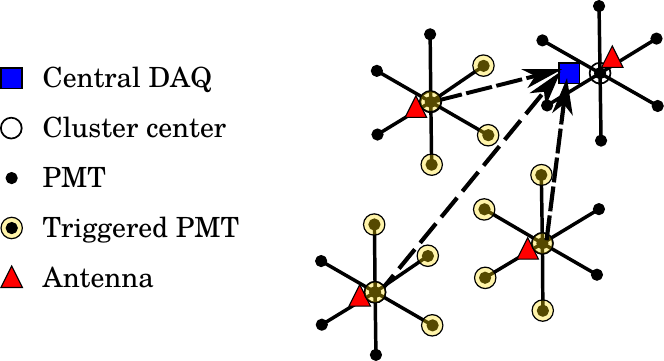}
\caption{Scheme of the data acquisition of Tunka-133/Tunka-Rex. Each cluster contains a local clock. The time for each single cluster event is calculated as cluster time plus the delay of the optical fiber. We assume that signals from the same air-shower are within a 7000~ns time window corresponding to the time needed by the shower front to cover the entire air-Cherenkov detector.} 
\label{fig_daq}
\end{figure}

Tunka-Rex is triggered by the photomultipliers and records the radio signal from the air showers between $30\,$ and $80\,$MHz, where the signal outside of this band is suppressed with an analog filter (Fig.~\ref{fig-spectrum}).
First, this improves the signal-to-noise ratio of the air-shower radio signals; second, this ensures that we measure entirely in the first Nyquist domain and can fully reconstruct the signal in this frequency band. 
Each of the clusters features its own local DAQ independently (for an air-shower event a coincidence in at least 3 PMTs is required). 
There the signal from both, the antennas and the PMTs, is digitized and transmitted to the central DAQ via optical fibers where it is stored on disk (see Fig.~\ref{fig_daq}).
Then, during the offline analysis, all independent entries from clusters are merged into hybrid events.

\begin{figure}[t!]
\centering
\includegraphics[width=1.0\linewidth]{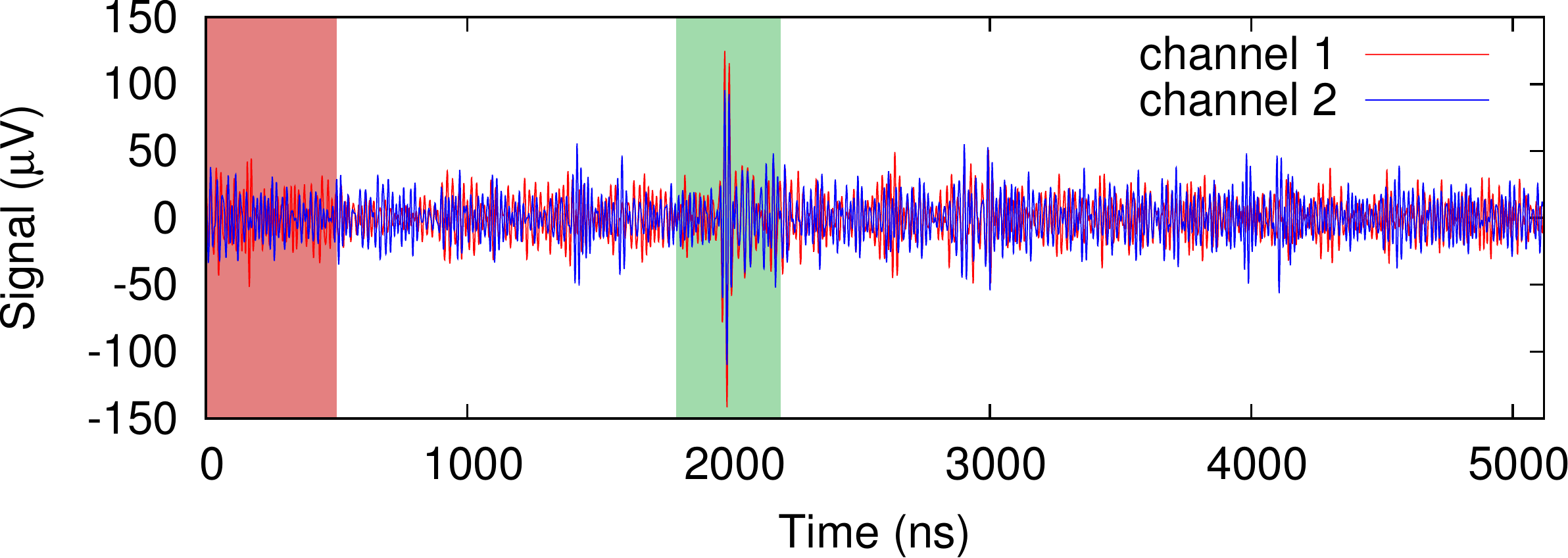}
\caption{Trace of the radio signal from air-showers before correcting for the antenna pattern. We take the first 500~ns (left shaded area) for the noise level estimate and estimate the bounds for the signal window (central shaded area) from cable lengths, hardware delays and detector geometry.} 
\label{fig_signal}
\end{figure}

Based on the known Tunka-133 and Tunka-Rex hardware properties, particularly, on the differences of cable lengths, we can estimate the window for the radio signal (see Fig.~\ref{fig_signal})
\begin{equation}
t = \frac{N_{FADC}}{2}T_{BW} + \frac{L_R - L_C}{\varepsilon c} - \frac{d}{\sqrt{2}c} -\tau \pm \Delta t\,,
\end{equation}
where $N_{FADC} = 1024$ is the number of FADC records in the trace, $T_{BW} = 5$~ns is the binwidth, $L_R \approx 30$ m, $L_C \approx 90$ m are the cable lengths to radio antennas and PMTs containers respectively, $c \approx 3\cdot 10^8$ m/s the speed of light, $\varepsilon c \approx (2/3)c$ is the signal velocity in the coaxial cable, $d \approx 80$ m is the typical distance between each cluster center and the surrounding PMTs, $\tau \approx 50$ ns is the PMT signal width and $\Delta t \approx d/(\sqrt{2}c) = 200$ ns is the bound for the radio signal window. The first term in this expression just gives the center of the trace (approximate position of the signal from last PMT), the second term is the delay due to the difference of cable lengths, the third term are possible delays due to geometry of the detector (for a typical zenith angle of 45$^\circ$), and the fourth term describes possible delay between shower arrival and triggering (passing amplitude threshold for PMT). Finally we have chosen $\Delta t$ to take into account all possible geometries (i.e. for the range of zenith angles from $0^\circ$ to $70^\circ$) of air showers. Thus, our estimation for the time window of the radio signal is $2000\pm200$ ns.

For more details on the Tunka-Rex hardware and the systematic uncertainties on signal reconstruction, see Ref.~\cite{HillerTunkaRex_ICRC2013}.

\section{Event selection and data analysis}

Tunka-Rex started operation on 8 October 2012. 
Since then operates within the Tunka-133 trigger, i.e. in dark moonless nights with good weather excluding the summer months from May to September. 

By design, the maximum zenith angle for each PMT illumination is 50$^\circ$ (PMTs are placed inside isolated metal barrels whose top is covered by plexiglas). 
The zenith angle for triggering can be extended up to 70$^\circ$ due to indirect detection of light reflected from the inner surfaces of barrels.
Thus, all radio events are divided in two groups based on geometry:
\begin{itemize}
\item "vertical" events: zenith angle $\theta < 50^\circ$ with reconstructed geometry, energy, shower maximum available from the air-Cherenkov detector.
These events are good candidates for the cross-calibration. The main disadvantage of them is the low statistics and small number of antennas per event due to the steepness  of the lateral distribution of the radio signal.
\item "horizontal" (inclined) events: zenith angle $\theta \ge 50^\circ$ with reliable reconstruction of the shower direction from the air-Cherenkov detector. For these events, the other shower parameters cannot be reconstructed from the air-Cherenkov measurements. Thus, if it is possible to reconstruct the shower parameters (energy, $X_{\mathbf{max}}$) from the radio measurements, this could increase the total statistics of usable events at Tunka.
\end{itemize}
For the exposure and flux estimation from the radio detector we still have to study the background in more details and make performance simulations.

\begin{figure}[b!]
\centering
\includegraphics[width=0.70\linewidth]{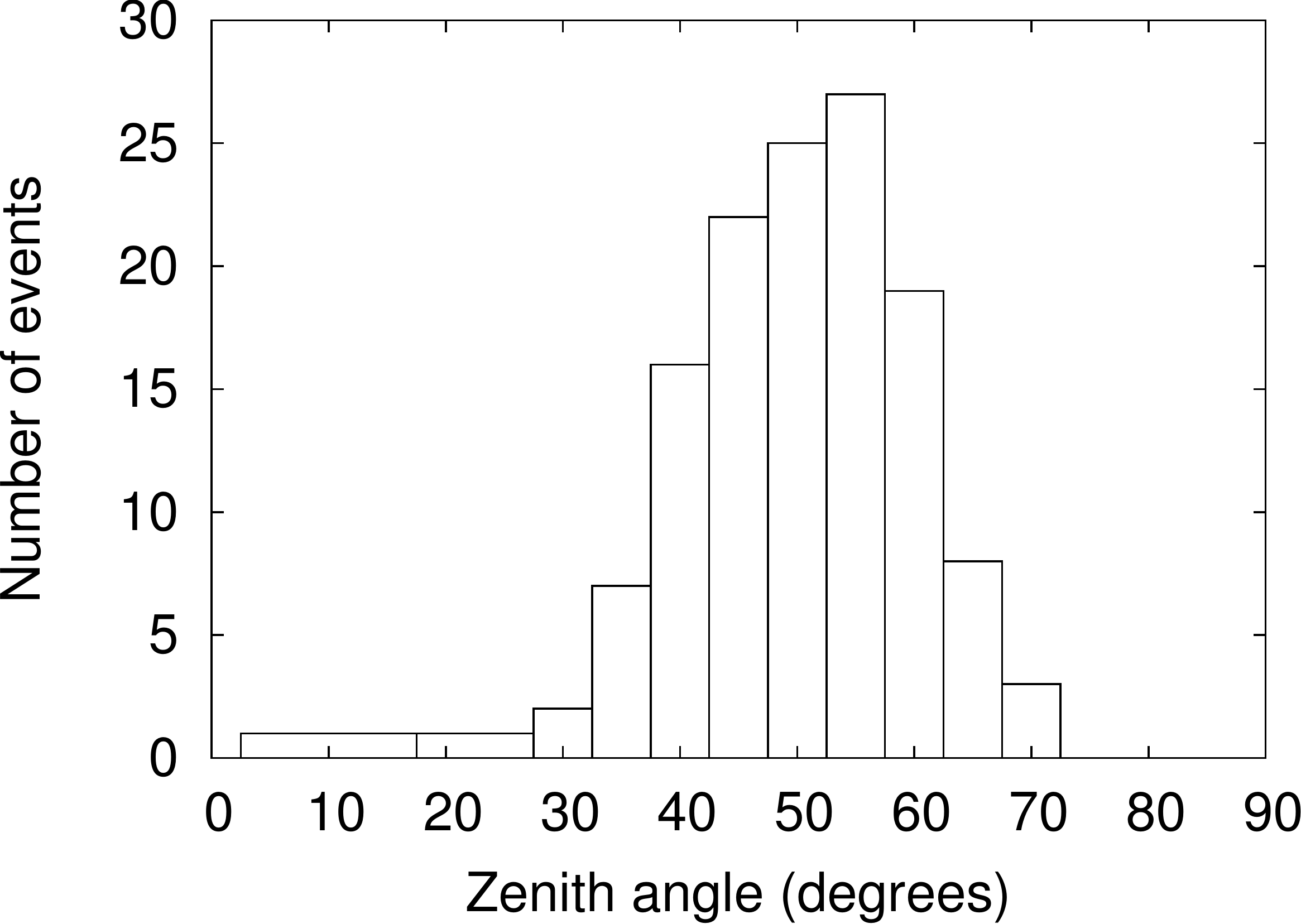}
\caption{Zenith angle distribution of radio detected events. The maximum efficiency is reached at the Tunka-133 reconstruction threshold of $\theta \approx 50^\circ$. The statistics at smaller angles is mainly suppressed by  the steeply falling lateral distribution of the radio signal, the statistics at larger angles is suppressed by trigger detection capabilities.} \label{fig_angle}
\end{figure}
\begin{figure*}[t!]
\centering
\includegraphics[width=0.85\linewidth]{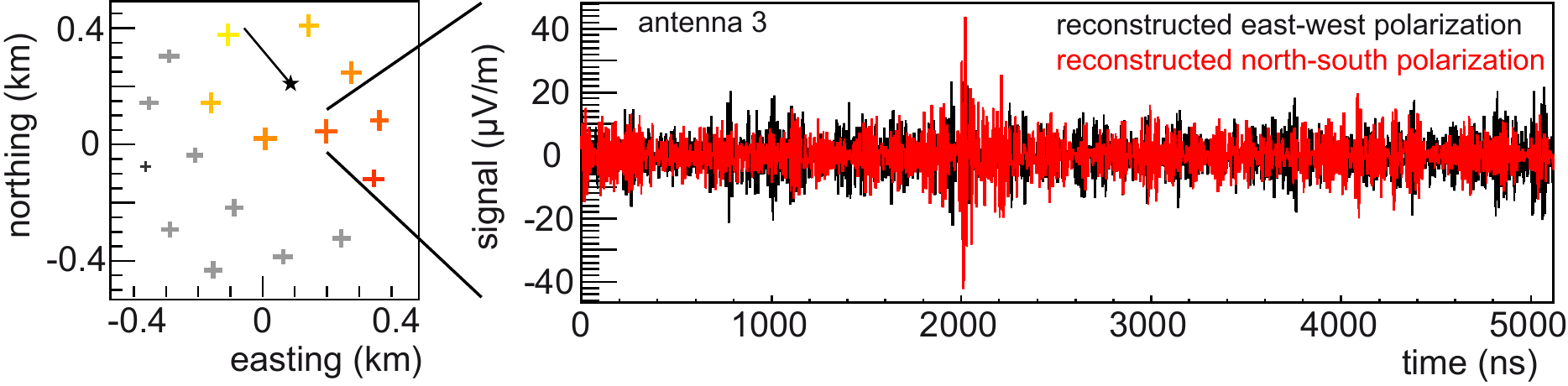}
\caption{Example for a Tunka-Rex event. Left: Footprint of the event, where the size of the crosses indicates the signal strength, the color code the arrival time, and the line and the star the direction and shower core respectively. Right: example trace of the reconstructed electric-field strength we expect the radio pulse around 2000~ns.} \label{fig_exampleEvent}
\end{figure*}
For a first analysis, we used only high quality events which have a clear radio signal (signal$^2$ / noise$^2$ $> 6$) in at least three antennas. Based on the detector specifications (opening angle for the PMT, typical distance between radio stations) one can assume that the maximum efficiency for hybrid events could be reached near the Tunka-133 reconstruction threshold $\theta \approx 50^\circ$ (Fig.~\ref{fig_angle}). Due to these reasons, only a small fraction of the air-Cherenkov events have also a clear radio signal (see Fig.~\ref{fig_exampleEvent} for an example).
Moreover we demand that the direction reconstructed from the arrival times of the radio signal agrees within $5^\circ$ with the direction obtained from the photomultiplier array. 
This cut excludes most of the background events, which by chance pass the signal-to-noise cut. 
In future, we plan to develop improved quality criteria based on the radio signal alone, to distinguish real from false events. 
For analysis of the radio measurements, we use the radio extension of the Offline software framework developed by the Pierre Auger Collaboration \cite{AugerOffline2007, RadioOffline2011}. 
It features a correction of the measured radio signal for the properties of the used hardware and a reconstruction of the electric field-strength vector at each antenna position. Since the absolute calibration of the antennas is still under evaluation we use a simulated pattern for the SALLA.


\begin{figure}[t]
\centering
\includegraphics[width=0.8\linewidth]{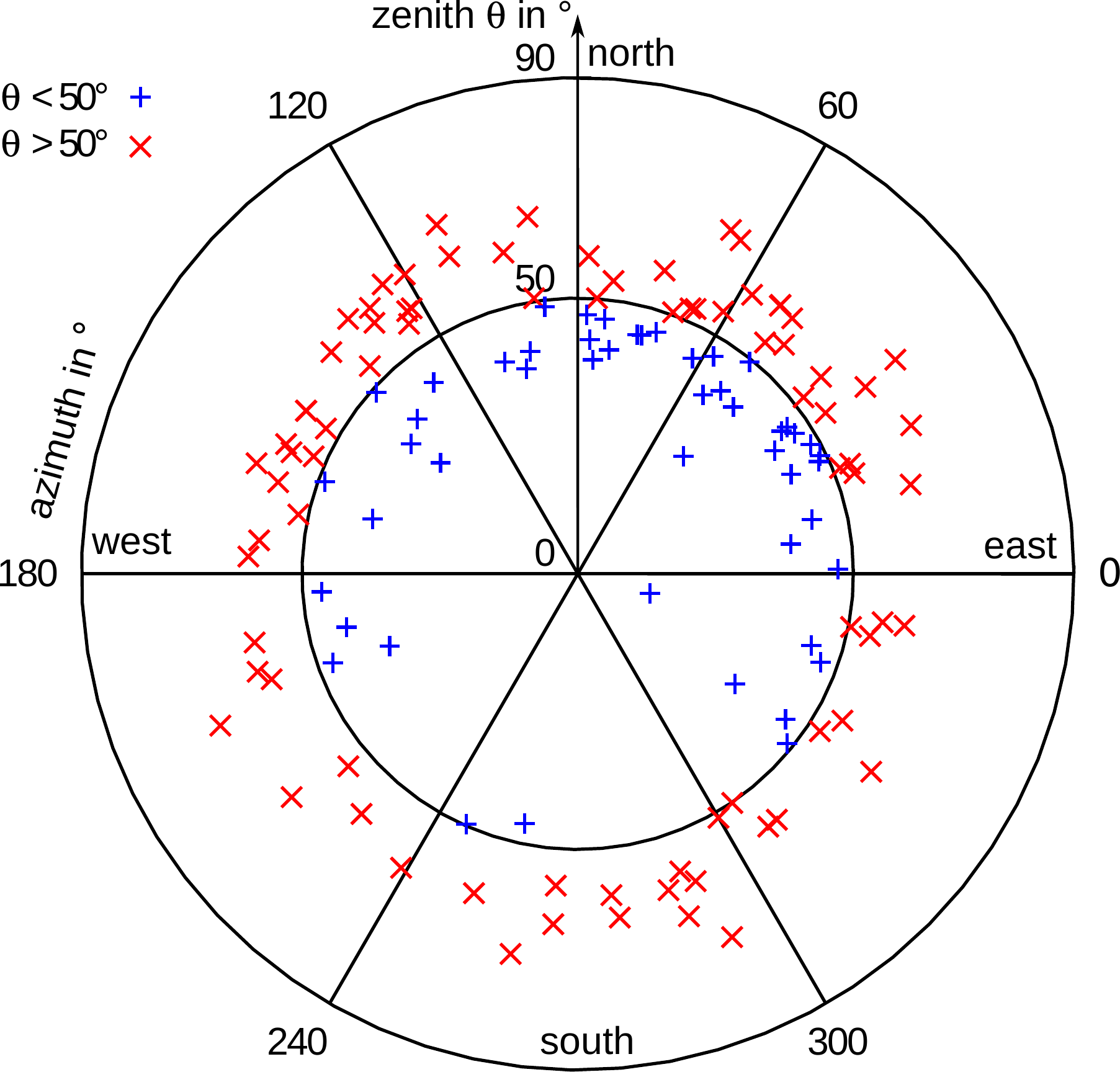}
\caption{Arrival directions of the Tunka-Rex events passing the quality cuts. Due to the geomagnetic effect, the radio signal is expected to be on average stronger for events coming from North, which explains the asymmetry in the detection efficiency: 89 of the 131 events are in the northern half of the plot, and 42 in the southern half.} 
\label{fig_angularDistribution}
\end{figure}

\begin{figure}[t]
\centering
\includegraphics[width=\linewidth]{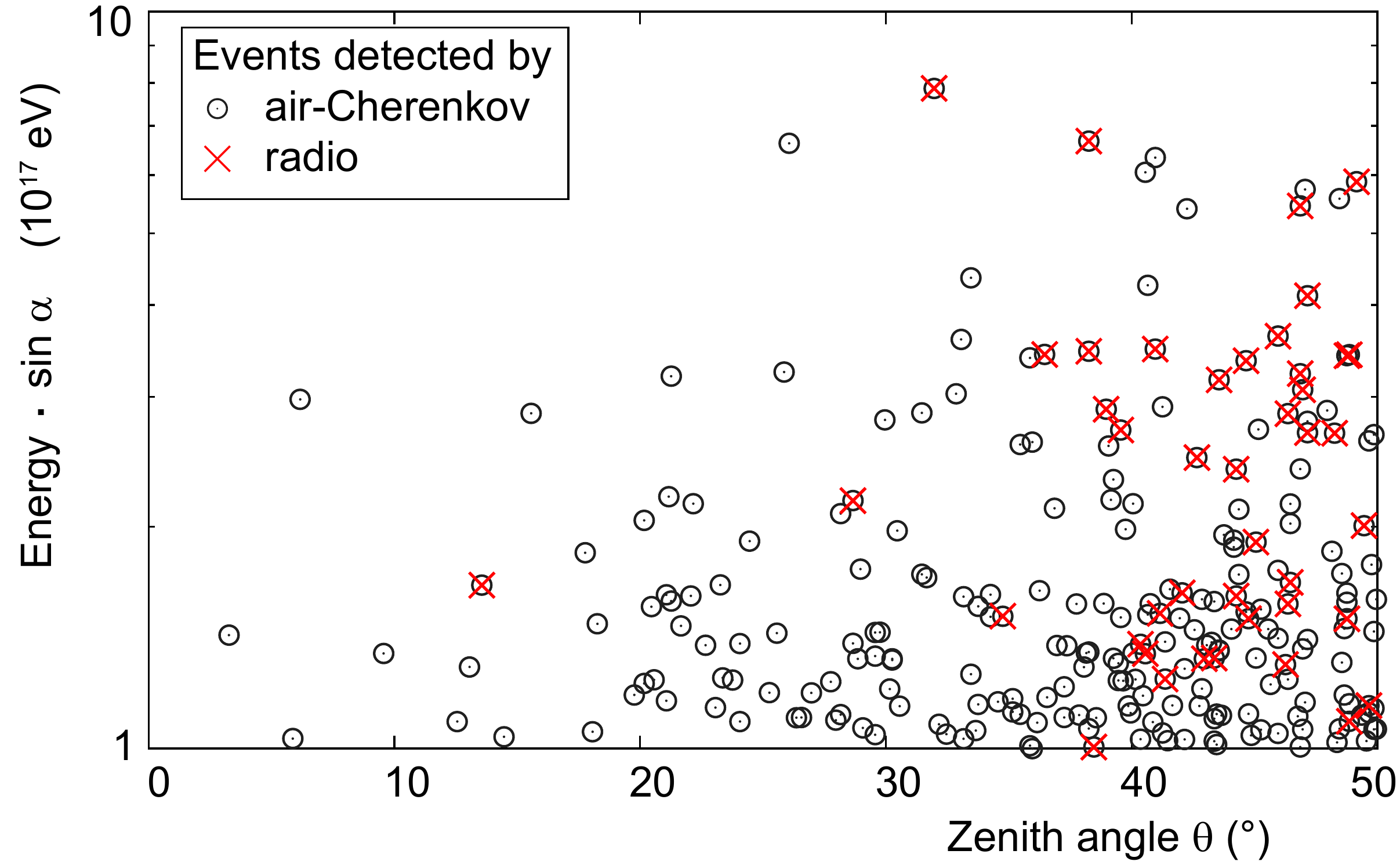}
\caption{Tunka events with energies above $10^{17}\,$eV detected by the PMT array, and those events passing the quality cuts for the radio measurements. The efficiency increases with energy $E$, the sine of the geomagnetic angle $\sin \alpha$, and the zenith angle $\theta$.} 
\label{fig_efficiency}
\end{figure}

\section{First Results}
Until now, we found 49 events with a zenith angle $\theta \le 50^\circ$, and 82 events with $\theta > 50^\circ$ in an effective measurement time of 392 hours (Table \ref{tab_EventStatistics}). 
Generally, the radio efficiency increases not only with large geomagnetic angles $\alpha$, i.e.~the angle between the shower axis and the geomagnetic field\footnote{The field strength $\varepsilon$ mainly depends on the vector product \\ $|\mathbf A \times \mathbf B| = |\mathbf A| \cdot |\mathbf B| \sin\alpha$, where $\mathbf A$ is the shower axis and $\mathbf B$ the geomagnetic field, $\alpha = \angle(\mathbf A, \mathbf B)$ is the geomagnetic angle.} (Figs.~\ref{fig_angularDistribution}~and~\ref{fig_efficiency}), but also with larger zenith angles (Fig.~\ref{fig_angle}). 
In addition, we expect that the event rate will increase when we complete the array to 25 antennas this autumn, and optimize our algorithms for digital background suppression.

\begin{table}[t!]
\centering
\caption{Statistics of Tunka-Rex events passing the quality cuts in dependence of the zenith angle $\theta$, excluding the period from 08 to 24 Oct 2012 used for commissioning of the detector. The effective measurement time and counting rate is limited by the PMT array, i.e.~light (moon) and weather conditions.} \label{tab_EventStatistics}
\vspace{0.1 cm}
\begin{tabular}{lccc} 
       &effective&\multicolumn{2}{c}{number of events}\\
measurement period &time& $\theta \le 50^\circ$ & $\theta > 50^\circ$\\
\hline
06 - 23 Nov 2012 & 56 h & 9 & 11 \\
04 - 23 Dec 2012 & 65 h & 8 & 12 \\
03 - 21 Jan 2013 & 114 h & 14 & 23 \\
01 - 17 Feb 2013 & 87 h & 12 & 22 \\
01 - 17 Mar 2013 & 70 h & 6 & 14 \\
\hline
Total sum & 392 h & 49 & 82 \\
\hline
\end{tabular}
\end{table}

\begin{figure}[t!]
\centering
\includegraphics[width=\linewidth]{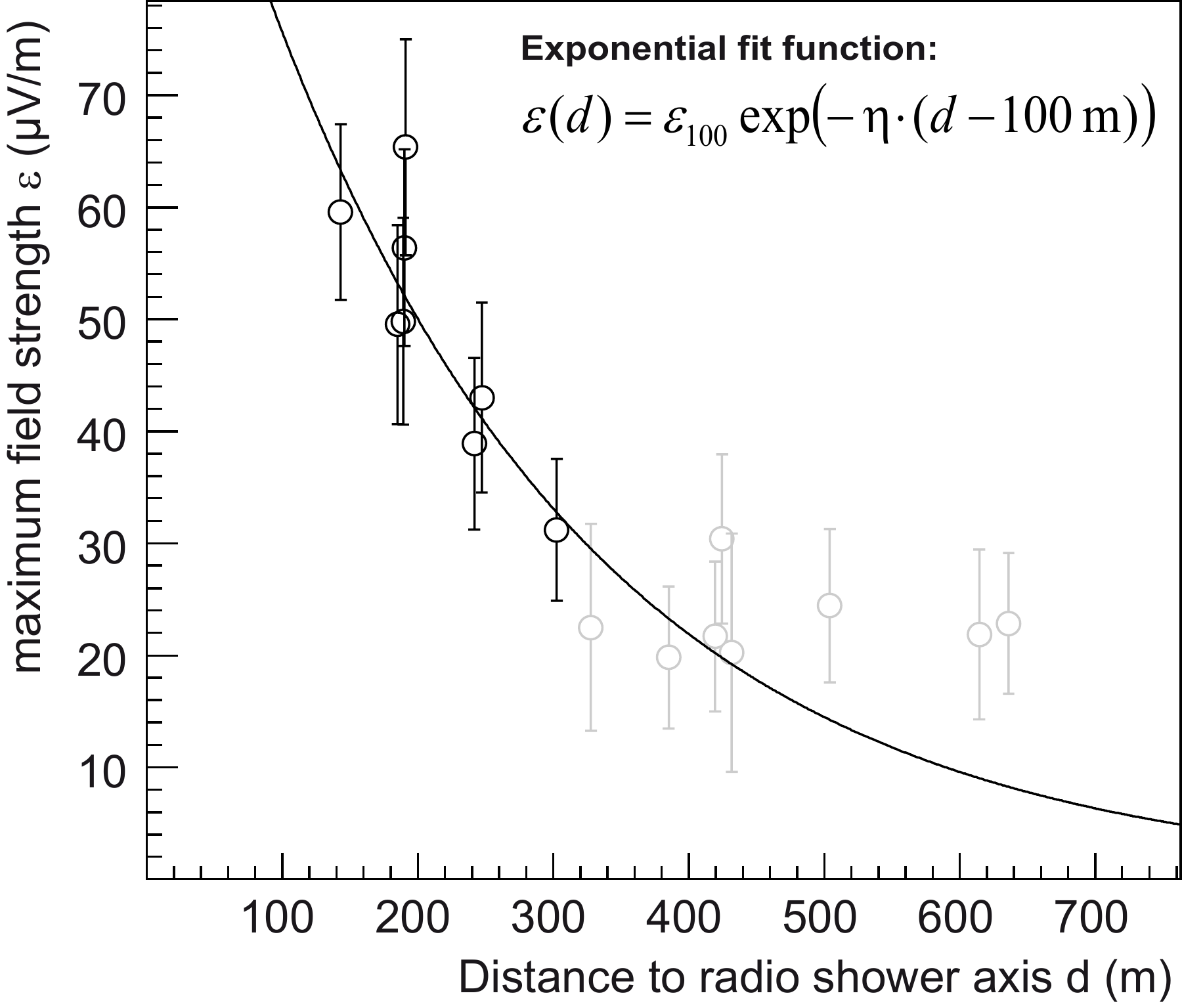}
\caption{Lateral distribution of the radio amplitude (i.e maximum field strength) for the example event shown in Fig.~\ref{fig_exampleEvent}. Light grey are antennas with signal-noise ratio below threshold (correspond to grey crosses in Fig.~\ref{fig_exampleEvent}).} 
\label{fig_exampleLDF}
\end{figure}

\begin{figure}[t!]
\centering
\includegraphics[width=\linewidth]{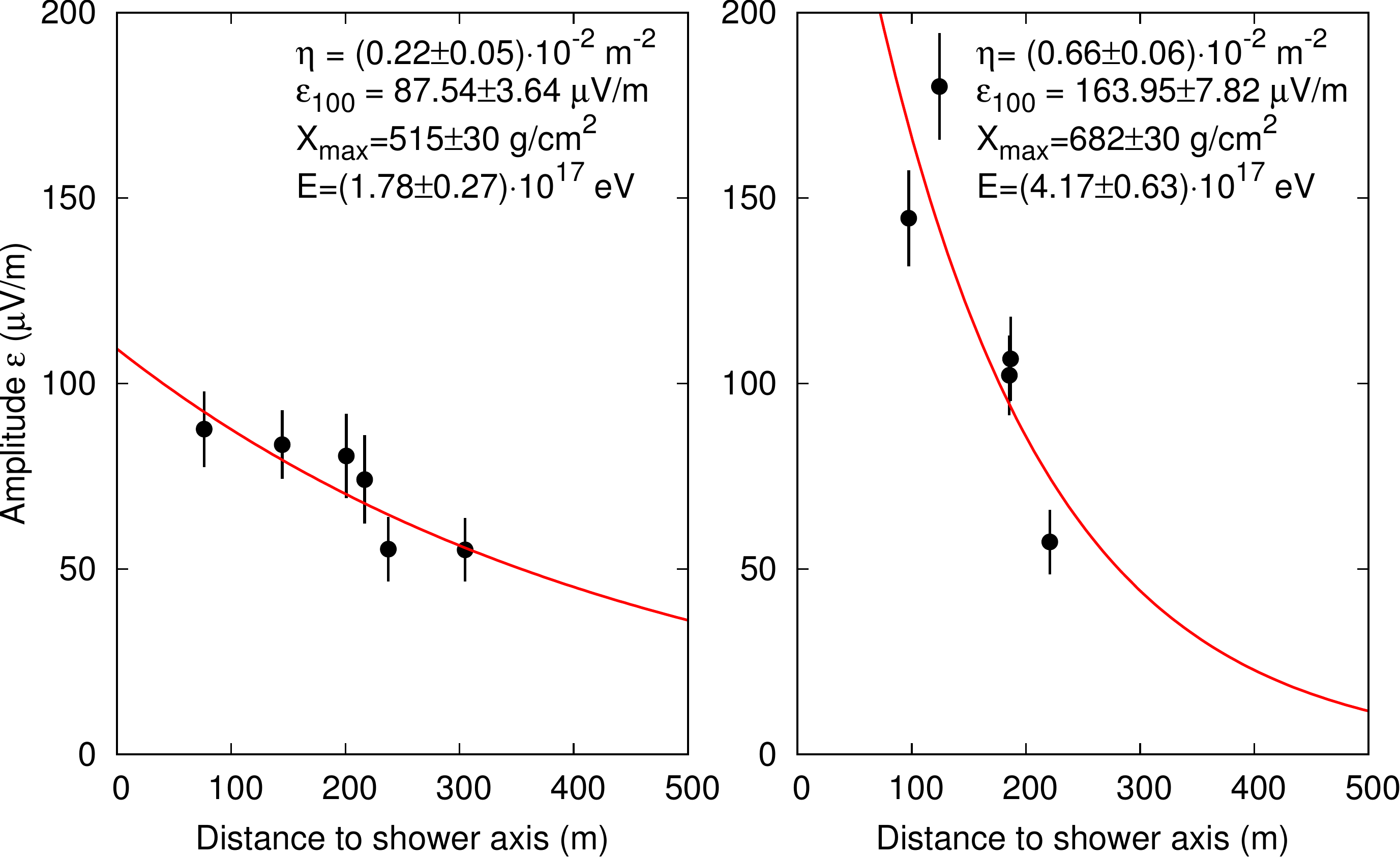}
\caption{Typical values of fitted parameters $\varepsilon_{100}$ (amplitude at 100 m), $\eta$ (slope parameter) in comparison with shower parameters obtained from the air-Cherenkov detector.} 
\label{fig_ldf2}
\end{figure}

To test the expected sensitivity of the Tunka-Rex measurements to air shower parameters, in particular to the energy, we reconstructed the lateral distribution of the radio signal for the 49 events with $\theta \le 50^\circ$. 
In a first approach, we used the shower geometry provided by the denser air-Cherenkov array to calculate the distance from each antenna to the shower axis, and then plotted the maximum absolute value of the electric field-strength vector as function of this axis distance. 
To estimate the uncertainties of the amplitude measurements and to correct the measured amplitudes for a bias due to background, we used formulas developed for the east-west aligned antennas of LOPES \cite{SchroederLOPESnoise_ARENA2010}, and then fitted an exponential function (Fig.~\ref{fig_exampleLDF}~and~\ref{fig_ldf2})\footnote{The exponential fit function has chosen according to the pioneer LOPES and CODALEMA experiments. First, this simple approximation was sufficient for the precision reached on these experiments, second, two parameters in that fit function could be easily connected to shower parameters. By the ongoing experiments and simulations it could be shown that the lateral distribution is more complicated and contains an azimuthal asymmetry due to the interference of the geomagnetic and the Askaryan effect.}.

\begin{figure}[t!]
\centering
\includegraphics[width=\linewidth]{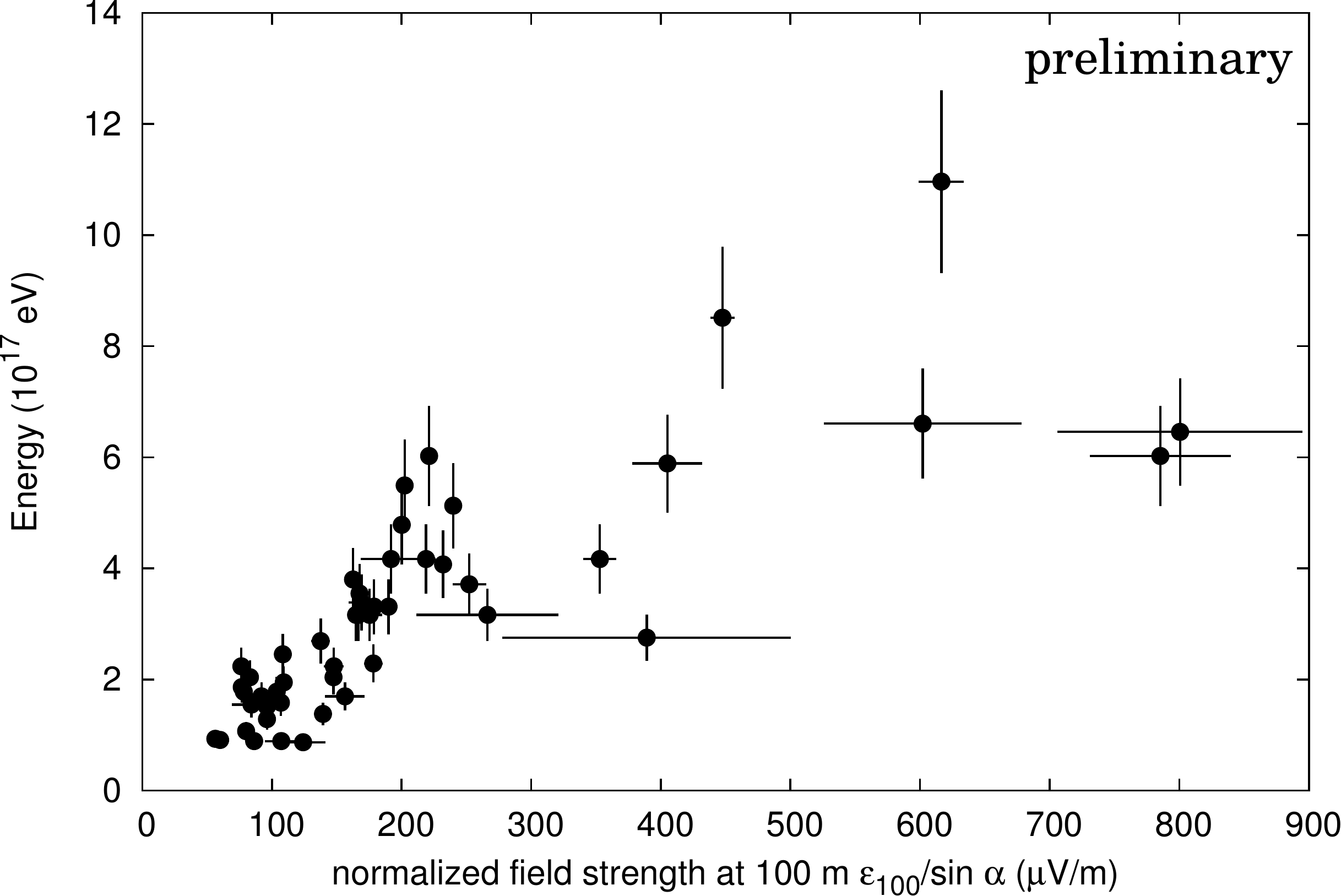}
\caption{Correlation between the radio field strength at 100 m normalized by sine of geomagnetic angle and the energy reconstructed with the air-Cherenkov measurements. The given values for the field strength are based on a simple exponential fit function and preliminary calibration.} 
\label{fig_energy}
\end{figure}

Consistent with several historic and modern experiments \cite{Allan1971, SchroederLOPES_ARENA2012, GlaserAERA_ARENA2012, RebaiCODALEMAenergy2012}, the amplitude parameter of the lateral distribution is correlated with the primary energy (Fig.~\ref{fig_energy}). However, the analysis is still preliminary, e.g., because of the status of the calibration and because the impact of the background at the Tunka site has to be studied in more detail. Moreover, we expect that the slope of the lateral distribution is sensitive to the position of the shower maximum \cite{HuegeUlrichEngel2008, deVries2010}, which we will analyze in near future by comparing Tunka-Rex measurements to the $X_\mathrm{max}$ reconstruction of the PMT array Tunka-133. Future work will be dedicated to find an optimal reconstruction algorithm for the energy and $X_\mathrm{max}$, and to test the achievable precision by comparison to the air-Cherenkov measurements. In addition to using the lateral distribution, $X_\mathrm{max}$ might also be obtained form the radio measurements via the shape of the radio wavefront \cite{Lafebre2010, SchroederWavefrontICRC2011}, or the slope of the frequency spectrum \cite{GrebeAERA_ARENA2012}.

\section{Conclusion}

As a result of the first weeks of operation, Tunka-Rex registered more than hundred events with significant radio signal from extensive air showers with energies above $10^{17}\,$eV in combination with the Tunka air-Cherenkov array. This shows that the Tunka observatory is able to provide hybrid measurements which is the pre-requisite to perform a cross-calibration between the air-Cherenkov and the radio signal.
Our measurements are compatible with the picture that the radio emission originates dominantly from the geomagnetic deflection of the charged particles in the air shower.

In future, we plan to optimize the reconstruction techniques, and to compare our measurements to simulations and other experiments.
A detailed background study might help to improve the signal-to-noise ratio and to increase the precision of the reconstruction of the shower maximum.
Moreover, we plan to trigger Tunka-Rex also by the planned scintillator extension of Tunka \cite{TunkaICRC2013}, and thus can measure also during day to increase the duty cycle by an order of magnitude. Finally, we will test a joint operation with Tunka-HiSCORE \cite{HiSCORE_RICAP2013,HiSCORE_ICRC2013} by deploying additional antennas. By this, we can also study to which extent a denser array of radio detectors can increase the detection efficiency and precision for the energy and $X_{\mathbf{max}}$ reconstruction.

\section*{Acknowledgments}
We acknowledge the support of the Russian Federation \\ Ministry of Education and Science (G/C 14.518.11.7046, \\ 14.B25.31.0010, 14.14.B37.21.0785, 14.B37.21.1294), the Russian Foundation for Basic Research (Grants 11-02-00409, 13-02-00214, 13-02-12095, 13-02-10001), the Helmholtz association (grant HRJRG-303).






\begin{thebibliography}{00}
\bibitem{Greisen:1966jv} 
  K.~Greisen,
  Phys.\ Rev.\ Lett.\  {\bf 16}, 748 (1966).
  
  \bibitem{Zatsepin:1966jv} 
  G.~T.~Zatsepin and V.~A.~Kuzmin,
  JETP Lett.\  {\bf 4}, 78 (1966)
  [Pisma Zh.\ Eksp.\ Teor.\ Fiz.\  {\bf 4}, 114 (1966)].

  
\bibitem{Askaryan1961}G.~A.~Askaryan, Sov. Phys. JETP \textbf{14}, 441 (1961)

\bibitem{Kahn1966}F.~D.~Kahn and I.~Lerche, Proc. Phys. Soc., Sect. A \textbf{289}, 206 (1966)

\bibitem{Castagnoli1969}C.~Castagnoli, G.~Silvestro, P.~Picchi, and G.~Verri, Nuovo Cimento B \textbf{63}, 373 (1969)

\bibitem{Hough1973} J.~H.~Hough, J.~Phys. A \textbf{6}, 892 (1973)

\bibitem{Jelley} J.~V.~Jelley, J.~H.~Fruin, N.~A.~Porter, \etal, Nature \textbf{205}, 327 (1965)

\bibitem{Vernov} S.~N.~Vernov, G.~B.~Khristiansen, A.~T.~Abrosimov, \etal, in Proceedings on the 11th ICRC, Budapest, Hungary (1969)

\bibitem{Allan1971}H.~R.~Allan, Prog. in Elem. Part. and Cosmic Ray Phys. \textbf{10} (1971) 171.

\bibitem{FalckeNature2005}H.~Falcke, \etal (LOPES Collaboration), Nature \textbf{435} (2005) 313.

\bibitem{CODALEMA}O.~Ravel for the CODALEMA Collaboration, NIM A \textbf{662} (2012) S89-S94

\bibitem{AERA}J.~Maller, these proceedings.

\bibitem{LOFAR}S.~Thoudam, these proceedings.

\bibitem{SchroederLOPES_ARENA2012}F.~G.~Schr\"oder, \etal (LOPES Collaboration), Proc. 5th ARENA, Erlangen, Germany, AIP Conf. Proc. \textbf{1535} (2013) 78.

\bibitem{GlaserAERA_ARENA2012}C.~Glaser, for the Pierre Auger Collaboration, Proc. 5th ARENA, Erlangen, Germany, AIP Conf. Proc. \textbf{1535} (2013) 68.

\bibitem{RebaiCODALEMAenergy2012}A.~Rebai, \etal (CODALEMA Collaboration), arXiv.org (2012) 1210.1739.

\bibitem{ApelLOPES_MTD2012}W.~D.~Apel, \etal (LOPES Collaboration), Phys. Rev. D \textbf{85} (2012) 071101(R).

\bibitem{PalmieriLOPES_ICRC2013}N.~Palmieri, \etal (LOPES Collaboration), Proc. 33rd ICRC, paper 0439, Rio de Janeiro, Brazil (2013)

\bibitem{LOFAR_ICRC2013} S.~Buitink, \etal (LOFAR Collaboration) Proc. 33rd ICRC, paper 0579, Rio de Janeiro, Brazil (2013)

\bibitem{TunkaRICAP2013} V.~Prosin, these proceedings.
\bibitem{TunkaICRC2013}N.~Budnev, for the Tunka Collaboration, Proc. 33rd ICRC, paper 0418, Rio de Janeiro, Brazil (2013)

\bibitem{AERAantennaPaper2012}The Pierre Auger Collaboration, JINST \textbf{7} (2012) P10011.
\bibitem{polarization}D.~Ardouin \etal - CODALEMA Collaboration, Astroparticle Physics, 31(3):192 – 200, 2009.

\bibitem{HillerTunkaRex_ICRC2013}R.~Hiller, \etal (Tunka-Rex Collaboration), Proc. 33rd ICRC, paper 1278, Rio de Janeiro, Brazil (2013)

\bibitem{AugerOffline2007}S.~Argiro, \etal, Nucl. Instr. Meth. A \textbf{580} (2007) 1485.

\bibitem{RadioOffline2011}The Pierre Auger Collaboration, Nucl. Instr. Meth. A \textbf{635} (2011) 92.

\bibitem{SchroederLOPESnoise_ARENA2010}F.~G.~Schr\"oder, \etal (LOPES Collaboration), Nucl. Instr. Meth. A \textbf{662} (2012) S238.

\bibitem{HuegeUlrichEngel2008}T.~Huege, R.~Ulrich, R.~Engel, Astropart. Phys. \textbf{30} (2008) 96.

\bibitem{deVries2010}K.~D.~de Vries, \etal, Astropart. Phys. \textbf{34} (2010) 267.

\bibitem{Lafebre2010}S. Lafebre, \etal, Astropart. Phys. \textbf{34} (2010) 12.

\bibitem{SchroederWavefrontICRC2011}F.~G.~Schr\"oder, \etal (LOPES Collaboration), Proc. 32nd ICRC Beijing, China (2011) \#0313.

\bibitem{GrebeAERA_ARENA2012}S.~Grebe, for the Pierre Auger Collaboration, Proc. 5th ARENA, Erlangen, Germany, AIP Conf. Proc. \textbf{1535} (2013) 73.

\bibitem{HiSCORE_RICAP2013}R.~Wischnewski, \etal (HiSCORE Collaboration), these proceedings.
\bibitem{HiSCORE_ICRC2013}R.~Wischnewski, \etal (HiSCORE Collaboration), Proc. 33rd ICRC, paper 1164, Rio de Janeiro, Brazil (2013)



\end{thebibliography}



\end{document}